\documentclass[12pt]{JHEP3}

\usepackage{epsfig}

\title{
The 3-D O(4) universality class
and the phase transition in two-flavor QCD}

\author{Francesco Parisen Toldin\\
        Scuola Normale Superiore \\ 
        Piazza dei Cavalieri 7, I-56126 Pisa, Italy.\\
	E-mail: \email{f.parisentoldin@sns.it} 
} 

\author{Andrea Pelissetto\\ 
	Dipartimento di Fisica dell'Universit\`a 
	di Roma  ``La Sapienza'' and INFN \\ 
        Piazzale Moro 2, I-00185 Roma, Italy \\
	E-mail: \email{Andrea.Pelissetto@roma1.infn.it}
}

\author{Ettore Vicari \\ 
	Dipartimento di Fisica dell'Universit\`a 
	di Pisa and INFN \\
        Via Buonarroti 2, I-56127 Pisa, Italy \\ 
	E-mail: \email{vicari@df.unipi.it} 
}

\abstract{
We determine the critical equation of state of the three-dimensional 
O(4) universality class.
We first consider the small-field expansion of 
the effective potential (Helmholtz free energy).
Then, we apply a systematic approximation scheme based
on polynomial parametric representations
that are valid in the whole critical regime,
satisfy the correct analytic properties (Griffiths' analyticity), 
take into account the Goldstone singularities at the coexistence curve,
and match the small-field expansion of the effective potential.
From the approximate representations of the  equation of state,
we obtain estimates of several universal amplitude ratios.

The three-dimensional O(4) universality class 
is expected to describe the finite-temperature chiral transition
of quantum chromodynamics with two light flavors. 
Within this picture,
the O(4) critical equation of state relates the
reduced temperature, the quark masses, and the
condensates around $T_c$ in the limit of vanishing quark masses.
}

\begin{document}

\section{Introduction}

In the theory of critical phenomena continuous phase transitions 
can be classified into universality 
classes determined only by a few properties characterizing the system, such
as the space dimensionality,  the range of interaction,
the number of components of the order parameter,
and the symmetry.  Renormalization-group theory predicts that 
critical exponents and scaling functions are 
the same for all systems belonging to a given universality class.
Here we consider the three-dimensional O(4) universality class, 
which is characterized by a four-component order parameter, O(4) symmetry,
and effective short-range interactions. 
A representative of this universality class is the lattice spin model
\begin{equation}
{\cal H} = - J \sum_{<ij>} \vec{s}_i\cdot \vec{s}_j - 
   \sum_i \vec{H}\cdot \vec{s}_i,
\label{spinmodel}
\end{equation}
where $\vec{s}_i$ are four-component unit spins
and the summation is extended over all nearest-neighbor pairs 
$<ij>$. We are interested in the critical 
equation of state that relates 
the magnetization  $\vec{M}\equiv {1\over V}\langle \sum_i \vec{s}_i\rangle$,
the temperature $T$, and a uniform external (magnetic) field $\vec{H}$
near the critical point $T=T_c$ and $\vec{H}=0$.
The three-dimensional O(4) model is relevant for the finite-temperature
behavior of quantum
chromodynamics (QCD) with two light-quark flavors.
Using symmetry arguments, it has been argued that,
if the finite-temperature transition is continuous, it should belong
to the same universality class of the three-dimensional O(4)
vector model (\ref{spinmodel}) \cite{PW-84,Wilczek-92,RW-93}. 

The thermodynamics of quarks and gluons described 
by QCD is characterized by a transition from 
a low-temperature hadronic phase, in which chiral symmetry
is broken, to a high-temperature phase with deconfined quarks and gluons
(quark-gluon plasma), in which chiral symmetry is restored.
The main features of this transition depend crucially on the QCD parameters, 
such as the number $N_f$ of flavors and the quark masses.
Our qualitative understanding of the deconfinement transition
is essentially based on the expected symmetry-breaking pattern 
that has been discussed at length in the light- 
\cite{PW-84,Wilczek-92,RW-93,GGP-94}
and in the heavy- \cite{SY-82} quark regime.
For recent reviews see, e.g., 
Refs.~\cite{Rajagopal-95,Wilczek-00,Karsch-01,Kogut-02,LP-03,Rischke-03}.
In the limit of infinitely heavy quarks, i.e. in the pure
SU(3) gauge theory, the order parameter for the transition is the Polyakov loop.
The corresponding effective theory turns out to be a three-dimensional 
$Z_3$-symmetric spin model, which is expected to undergo a first-order 
transition. 
With decreasing the quark masses, 
the first-order transition should persist up to a critical surface
(a line if only two quarks are considered) in the quark-mass phase diagram, 
where the transition becomes continuous and is expected to be Ising-like 
\cite{PW-84}.
When the quark masses are further decreased, the phase transition 
disappears and one only observes an analytic crossover. 
In the opposite limit, i.e. for vanishing quark masses,
we expect again a phase transition that is essentially
related to the restoring of the chiral symmetry.
Its nature can be argued by using symmetry arguments. 
In the chiral limit, since the axial $U(1)_A$ symmetry is broken by the 
anomaly, the relevant global symmetry group is expected to be
$SU(N_f)_L\times SU(N_f)_R$.
At $T=0$ the symmetry is spontaneously
broken to $SU(N_f)_{L+R}$ with $N_f^2-1$ Goldstone particles (pions
and kaons).
With increasing $T$, QCD is expected to undergo 
a chiral symmetry-restoring transition, where
the order parameter is played by the expectation value of the quark bilinear
$M^i_j=\langle\bar{q}^i_L q_{Rj}\rangle$.
In the case the chiral symmetry-restoring phase transition is of first-order type, 
it persists also for
nonvanishing masses, up to a critical surface in the quark-mass phase diagram
where the transition becomes continuous and is expected to be Ising-like \cite{GGP-94}.
Outside this surface, i.e. for larger quark masses,
the phase transition disappears and we are back in the 
analytic-crossover region discussed above.
In the case the phase transition is continuous, 
its universal critical behavior can be  described
by an effective three-dimensional theory.\footnote{A 
finite-temperature $d$-dimensional quantum field theory is equivalent 
to a classical model defined in ($d$+1) dimensions  with a finite extent 
in the ``temporal" ($d$+1) dimension. If the transition is continuous, 
near the critical point the critical modes have a correlation length much 
larger than the temporal extension that can therefore be neglected. 
Thus, the theory that gives the universal features of the transition is  
effectively $d$-dimensional.}
Moreover, an analytic crossover is expected for
nonvanishing quark masses, since the quark masses act as a external field
coupled to the order parameter.

In the two-flavor case, $N_f=2$, the symmetry-breaking pattern of the 
transition is equivalent to $O(4)\rightarrow O(3)$. 
According to universality arguments, if two-flavor QCD 
undergoes a continuous transition, its critical behavior at $T_c$
should be that of the three-dimensional O(4) universality class. 
Since the quark masses act as an external (magnetic) field, a quark-mass term
smooths out the singularity and instead of the transition one observes 
an analytic crossover, as in the O(4) model for nonvanishing magnetic field. 
In the small quark-mass regime the universal features of this analytic 
crossover can be still described by the scaling O(4) theory.
However, universality arguments do not exclude that the transition is 
of first order. In this case, 
it would appear even for small nonvanishing masses.
A first-order transition should be expected if the breaking of the axial 
$U(1)_A$ symmetry is effectively weak at $T_c$. Indeed, 
in the absence of the $U(1)_A$ axial anomaly,
the symmetry breaking pattern is 
$U(2)_L\times U(2)_R \rightarrow U(2)_{L+R}$. The corresponding 
effective three-dimensional Landau-Ginzburg-Wilson
theory~\cite{PW-84} does not have stable fixed points,
and therefore one expects a first-order transition.\footnote{
This picture was originally argued on the basis
of a first-order perturbative calculation in the framework
of the $\epsilon\equiv 4-d$ expansion \cite{PW-84}.
Recently, a six-loop computation within a three-dimensional perturbative 
scheme \cite{BPV-03} put this result on a firmer ground. }
Since a first-order transition is generally robust
under perturbations, for a sufficiently small 
breaking of the axial $U(1)_A$ symmetry at $T_c$ 
the transition should maintain its first-order 
nature. This issue has been investigated 
on the lattice, see, e.g., Refs.~\cite{Betal-97,KLS-98,Vranas-99}.
The $U(1)_A$ symmetry is not restored at $T_c$ and
the transition is apparently of second order.\footnote{
The $U(1)_A$ symmetry appears not to be restored at $T_c$.
We mention that the effective breaking of the axial $U(1)_A$ symmetry 
appears substantially reduced especially above $T_c$,
as inferred from the difference between the correlators
in the pion and $\delta$ channels \cite{BCM-97,Betal-97,KLS-98,Vranas-99}.
}

In the case of three light flavors, $N_f=3$, 
on the basis of the symmetry arguments reported above,
it has also been argued that the transition is of first order, up to 
a critical surface in the quark-mass phase diagram, where it 
becomes continuous and it is expected to be in the Ising universality 
class \cite{GGP-94}. This picture has been
also supported by a recent Monte Carlo simulation \cite{KLS-01}.
See, e.g., Ref.~\cite{Karsch-01} for a discussion of the 
phase diagram for generic values of the quark masses. 
A similar behavior has been put forward for $N_f>3$ \cite{PW-84}.
\footnote{Again, this has been argued on the basis
of a first-order $\epsilon$-expansion perturbative calculation 
in the framework of the corresponding effective three-dimensional 
quantum field theory.
This has been recently confirmed by 
a six-loop computation within a three-dimensional perturbative 
scheme \cite{BPV-03}.} 

Since many years the finite-temperature behavior of QCD 
has been investigated by numerical Monte Carlo
simulations exploiting the lattice formulation of
QCD. The available results are substantially consistent
with the above-reported picture based on symmetry and universality, 
see, e.g., Refs. \cite{CP-PACS-01,IKKY-97,IKSY-96,KLP-01,KS-01,Betal-00}.
However, although the evidence for a continuous transition in
two-flavor QCD at $T_c\approx 172$ MeV is rather convincing
(see, e.g., Refs.~\cite{CP-PACS-01,KLP-01}),
conclusive evidence in favor of an O(4) scaling
behavior in the continuum limit has not been achieved yet.

An accurate determination of the universal features of
three-dimensional O(4) systems, such as the critical equation of state,
is important to achieve an unambigous identification of
the universality class of the finite-temperature transition in
two-flavor QCD. Moreover, 
assuming that two-flavor QCD undergoes a continuous
transition belonging to the three-dimensional O(4) universality
class, the O(4) critical equation of state gives the
asymptotic relations, for $T\rightarrow T_c$ and in the
limit of vanishing quark masses, among
the reduced temperature $t\equiv T/T_c-1$, the quark masses, and the
expectation value of the bilinear 
$M^i_j=\langle\bar{q}^i_L q_{Rj}\rangle$.
In particular, in
the two-flavor case  the relevant order parameter can be written
as a four-component vector $\vec{\varphi}$, whose expectation value  
$\vec{M}$  is related to the quark condensate
$\langle \bar{q} q \rangle$, and which is coupled to a
uniform external field $\vec{H}$ related to the quark masses 
\cite{PW-84,Wilczek-92,RW-93,Rajagopal-95}.
The scaling properties can be written as
\begin{eqnarray}
&&\vec{M} = B_c \vec{H} |H|^{(1-\delta)/\delta} E(y),
\label{eyf}\\
&&y=(B/B_c)^{1/\beta} t |H|^{-1/(\beta+\gamma)},
\nonumber
\end{eqnarray}
where $\beta$, $\gamma$, and $\delta$ are the 
critical exponents,\footnote{
The critical exponents of the O(4) universality class
have been determined rather accurately by exploiting
field-theoretical methods and lattice techniques.
We mention the field-theory estimates of Ref.~\cite{GZ-98},
$\nu=0.741(6)$ and $\eta=0.035(5)$, and
the Monte Carlo results of Ref.~\cite{Hasenbusch-00},
$\nu=0.749(2)$ and $\eta=0.0365(10)$.
A more complete list of references and results
can be found in Ref.~\cite{PV-r}.
The other exponents can be obtained by using scaling
and hyperscaling relations:
$\alpha=2-3\nu$,  $\gamma=\nu(2-\eta)$,
$\beta=\nu (1+\eta)/2$ and $\delta=(5-\eta)/(1+\eta)$.
For example, using 
the most precise Monte Carlo results one obtains
$\alpha=-0.247(6)$, $\gamma=1.471(4)$, $\beta=0.388(1)$, and $\delta=4.789(6)$.
} 
$B$ and $B_c$ are nonuniversal constants 
fixing the normalizations (they are the amplitudes of the magnetization
at the coexistence curve and along the critical isotherm, respectively),
and $E(y)$ is a universal function.
Similar asymptotic relations can also be written for the 
(quark-mass) susceptibilities; they can be easily derived by taking 
appropriate derivatives of the relation (\ref{eyf}).
Corrections to the scaling behavior (\ref{eyf})
are suppressed by powers of $t$ and $|H|$.
The leading ones are of order $t^\Delta$ on the axis $H=0$ 
and of order $|H|^{\Delta/(\beta+\gamma)}$ on the critical isotherm
[in general corrections have the scaling behavior $t^\Delta g(y)$],
where $\Delta\approx 0.58$ \cite{GZ-98,Hasenbusch-00}.

The critical equation of state of $O(N)$ models has already been 
studied in $\epsilon\equiv 4-d$ expansion at two
loops \cite{BWW-72} and in $1/N$ expansion at order $1/N$ \cite{BW-73}.
However, these results do not allow us to obtain  a quantitatively
accurate determination of the critical equation of 
state, essentially because the available expansions are too short.
On the other hand, on the numerical side,
rather accurate results have been obtained
from Monte Carlo simulations of the O(4) spin model (\ref{spinmodel}) 
\cite{Toussaint-97,EM-00,EHMS-01}.

In this paper we compute the critical equation of state
by using a different analytic method.
The starting point is the small-field expansion of the effective potential 
(in statistical-mechanics language it corresponds to 
the small-magnetization expansion of the Helmholtz
free energy), or equivalently of the equation of state,
in the symmetric (high-temperature) phase ($t>0$).
The first few nontrivial terms of this expansion have already been 
determined by exploiting field-theoretical methods 
in the continuum $\varphi^4$ theory \cite{PV-effpot,SOUK-99}.
We use them to determine  approximations of the equation of state 
that are valid in the whole critical regime.
This requires an analytic continuation from 
the high-temperature phase ($t>0$) to the the coexistence curve ($t<0$),
which can be achieved by using parametric
representations \cite{ZJbook,GZ-97,CPRV-99-02,CPRV-00} 
implementing in a rather simple  way the known analytic 
properties of the equation of state (Griffiths' analyticity).
We construct systematic approximation schemes based on
polynomial parametric representations
of the critical equation of state that have the correct analytic 
properties, take into account the Goldstone singularities at the
coexistence curve, and match the known small-field
behavior of the effective potential. 
Using the results for the critical equation of state,
we then derive estimates of several universal 
amplitude ratios.
As we shall see, we will provide rather accurate determinations of 
the scaling functions related to the magnetization, the longitudinal
(quark-mass) susceptibility, and of several universal amplitude
ratios, involving also quantitites defined at the so-called pseudo-critical 
line that corresponds to 
the maximum of the longitudinal susceptibility at fixed $H$.

The paper is organized as follows.
In Sec.~\ref{CES} we discuss the general properties of the
critical equation of state. In particular, we consider the
small-field expansion of the effective potential and of 
the critical equation of state in the symmetric phase.
In Sec.~\ref{preq} we describe our approximation schemes
based on polynomial parametric representations,
and determine the scaling functions that give the scaling behavior
of the magnetization (quark condensate) and of the longitudinal
susceptibility.
Finally, in Sec.~\ref{unratio} we determine several universal 
amplitude ratios.

\section{The critical equation of state}
\label{CES}

\subsection{General properties} 
\label{CES1}

The equation of state relates the 
magnetization $\vec{M}$, the magnetic field $\vec{H}$, and the reduced temperature 
$t\equiv (T-T_c)/T_c$. 
In the neighborhood of the critical point $t=0$, $H=0$, it is usually 
written in the scaling form 
\begin{equation}
 \vec{H} = B_c^{-\delta}  \vec{M} M^{\delta-1} f(x), \qquad
x \equiv t (M/B)^{-1/\beta},
\label{eqstfx}
\end{equation}
where $M\equiv |\vec{M}|$,
$f(x)$ is a universal scaling function normalized in such a way
that $f(-1)=0$, $f(0)=1$, and 
$B_c$ and $B$ are the amplitudes of the magnetization on the critical 
isotherm and at the coexistence curve:
\begin{eqnarray}
&&M=B_c H^{1/\delta}, \qquad t=0, \label{bc}\\
&&M=B (-t)^\beta, \qquad t<0,\quad H\rightarrow 0. \label{bt}
\end{eqnarray}
The scaling function $f(x)$ and $E(y)$, cf. Eq.~(\ref{eyf}),
are clearly related:
\begin{eqnarray}
E(y) = f(x)^{-1/\delta} , \qquad
y = x f(x)^{-1/(\beta + \gamma)}.
\label{rel-funw-funx}
\end{eqnarray}

The equation of state is analytic for $|H|>0$, and therefore
$f(x)$ is regular everywhere for $x>-1$. 
In particular, $f(x)$ has a regular expansion in powers of $x$
around $x=0$
\begin{equation}
f(x) = 1 + \sum_{n=1}^\infty f_n^0 x^n,
\label{expansionfx-xeq0}
\end{equation}
and a large-$x$ expansion of the form
\begin{equation}
f(x) = x^\gamma \sum_{n=0}^\infty f_n^\infty x^{-2n\beta}.
\label{largexfx}
\end{equation}

At the coexistence curve, i.e. for $x\rightarrow -1$,
the Goldstone singularities appear.
General arguments predict that, at the
coexistence curve and in three dimensions,
transverse and longitudinal 
susceptibilities behave respectively as
\begin{equation}
\chi_T = {M\over H}  ,\qquad\qquad
\chi_L = {\partial M\over \partial H} \sim H^{-1/2}.
\label{chitl}
\end{equation}
In particular, the singularity of $\chi_L$ for $t<0$ and $H\to0$ is
governed by the zero-temperature infrared-stable Gaussian fixed point
\cite{BW-73,BZ-76,Lawrie-81}, leading to the prediction
\begin{equation}
f(x) \approx  f_2^{\rm coex} \,(1+x)^2 \qquad\qquad {\rm for}
\qquad x\rightarrow -1.
\label{fxcc} 
\end{equation}
The nature of the corrections to the behavior (\ref{fxcc}) is less
clear.  It has been conjectured \cite{WZ-75,SH-78,Lawrie-81}, using
essentially $\epsilon$-expansion arguments, that, for $x\to -1$, i.e.,
near the coexistence curve, $v\equiv 1+x$ has a double expansion in
powers of $w\equiv H M^{-\delta}$ and $w^{(d-2)/2}$.  This would imply
that in three dimensions $f(x)$ could be expanded in integer powers of $v$ at
the coexistence curve.  On the other hand, an explicit calculation
\cite{PV-99} to next-to-leading order in the $1/N$ expansion shows the
presence of logarithms in the asymptotic expansion of $f(x)$ for
$x\rightarrow -1$, so that $f(x)$ cannot be expanded in powers of $v$.  
However, these nonanalytic terms represent small corrections of order
$O(v^2 \log v)$ compared to the leading behavior (\ref{fxcc}).

Finally, beside the scaling functions $E(y)$ and $f(x)$,
we introduce a scaling function associated with the 
longitudinal susceptibility, by writing 
\begin{equation}
\chi_L = B_c |H|^{1/\delta-1} D(y),
\label{defDy-1}
\end{equation}
where 
\begin{eqnarray}
D(y) = {1\over \delta} 
   \left[E(y) - {y\over \beta} E'(y)\right] 
    = {\beta f(x)^{1-1/\delta} \over 
         \beta\delta f(x) - x f'(x)}.
\label{defDy-2}
\end{eqnarray}
The function $D(y)$ has a maximum at $y=y_{\rm max}$ corresponding to 
the crossover or pseudocritical line $t_{\rm max}(H)$,
see also Sec.~\ref{unratio}.

\subsection{The small-field expansion of the effective potential}
\label{CES2}

Let us consider the continuum $\varphi^4$ theory that describes
the three-dimensional O(4) universality class, i.e.
\begin{equation}
S = \int d^3x \Bigl\{ {1\over 2}
    \partial_\mu \vec{\varphi}(x)\cdot \partial_\mu \vec{\varphi}(x) + 
   {1\over 2} {r} \,\vec{\varphi}(x)\cdot \vec{\varphi}(x)  
      +{1\over 4!}u \left[ \vec{\varphi}(x)\cdot \vec{\varphi}(x) \right]^2 
       - \vec{H} \cdot\vec{\varphi}(x)\Bigr\},
\end{equation}
where $\vec{\varphi}(x)$ is a four-component real field.
Let us also consider a zero-momentum renormalization scheme \cite{Parisi-80}
(see also Refs.~\cite{ZJbook,It-Dr-book}):
\begin{eqnarray}
&&\Gamma^{(2)}_{ab}(p) = \delta_{ab} Z_\varphi^{-1} \left[ m^2+p^2+O(p^4)\right],
\label{ren1} \\
&&\Gamma^{(4)}_{abcd}(0) = 
Z_\varphi^{-2} m
{g\over 3}\left(\delta_{ab}\delta_{cd} + \delta_{ac}\delta_{bd} +
                \delta_{ad}\delta_{bc} \right),
\label{ren2}
\end{eqnarray}
where $\Gamma^{(n)}_{a_1,\ldots,a_n}$ are $n$-point one-particle irreducible 
correlation functions, $m$ and $g$ are the zero-momentum mass scale and
quartic coupling, respectively.

The effective potential (Helmholtz free energy) is related to the
(Gibbs) free energy of the model.  If $\vec{M}\equiv\langle
\vec{\varphi}\rangle$ one  defines
\begin{equation}
{\cal F} (M) = \vec{M} \cdot \vec{H} - {1\over V} \log Z(H),
\end{equation}
where $Z(H)$ is the partition function and the dependence on the
temperature is always understood in the notation.
The effective potential ${\cal F} (M)$ is 
the generator of the zero-momentum one-particle irreducible correlation 
functions.  In the high-temperature phase it admits an expansion around $M=0$:
\begin{equation}
\Delta {\cal F} \equiv {\cal F} (M) - {\cal F} (0) = 
\sum_{j=1}^\infty {1\over (2j)!} a_{2j} M^{2j}.
\end{equation}
This expansion can be rewritten in terms of the expectation value of the 
renormalized field  $\varphi_r(x) = Z_\varphi^{-1/2} \varphi(x)$, i.e.
$\vec{M}_r\equiv \langle \vec{\varphi}_r \rangle$,
\begin{equation}
\Delta {\cal F}= {1\over 2} m^2 M_r^2 + 
\sum_{j=2} m^{3-j} {1\over (2j)!} g_{2j} M_r^{2j}.
\label{freeeng}
\end{equation}
Note that $g_4$  is the renormalized coupling $g$ 
appearing in Eq.~(\ref{ren2}); its critical limit is
the fixed point of the renormalization-group equations, 
that is the zero of the Callan-Symanzik $\beta$-function 
$\beta(g)\equiv m\partial g/\partial m$.
The coefficients  $g_{2j}$ approach universal
constants (which we indicate with the same symbol) for $m\to 0$.
By performing a further rescaling
\begin{equation}
M_r^2 = {m \over g_4} z^2
\label{defzeta}
\end{equation} 
in Eq.\ (\ref{freeeng}), 
the effective potential can be  written as
\begin{equation}
\Delta {\cal F} = {m^3\over g_4}A(z),
\label{dAZ}
\end{equation}
where
\begin{equation}
A(z) =   {1\over 2} z^2 + {1\over 4!} z^4 + \sum_{j=3} {1\over (2j)!} r_{2j} z^{2j},
\label{AZ}
\end{equation}
and
\begin{equation}
r_{2j} = {g_{2j}\over g_4^{j-1}} \qquad\qquad j\geq 3.
\label{r2j}
\end{equation}
The universal quantities $g_4$ and the first few $r_{2j}$
have been estimated by using field-theoretical methods. 
Fixed-dimension perturbative calculations provide the estimates
$g_4=17.30(6)$ (from an analysis \cite{GZ-98} of the six-loop expansion of
$\beta(g)$ \cite{BNGM-77}),
$r_6=1.81(3)$ (from an analysis \cite{PV-effpot} of the corresponding 
four-loop series \cite{SOUK-99})
and $r_8=0.456$ (three-loop series \cite{SOUK-99}),
while $\epsilon$-expansion computations give \cite{PV-effpot}
$g_4=17.5(3)$ (four-loop series), 
$r_6=1.780(8)$ and $r_8=0.2(4)$ (three-loop series).
Other substantially less precise results 
and references can be found in Refs.~\cite{PV-effpot}.

Since $z\propto t^{-\beta} M$ where $t\equiv T/T_c-1\propto r-r_c$, 
the equation of state can be written in the form
\begin{equation}
\vec{H} = {\partial {\cal F}(M)\over \partial \vec{M}} \propto 
{\vec{M}\over |M|} t^{\beta\delta} F(z),
\label{eqa}
\end{equation}
with 
\begin{equation}
F(z) \equiv {\partial A \over \partial z}=
z + {1\over 6} z^3 + 
\sum_{j=3} {1\over (2j-1)!} r_{2j} z^{2j-1}.
\label{fzexp}
\end{equation}
The two functions $f(x)$ and $F(z)$ are related:
\begin{equation}
z^{-\delta} F(z) = F_0^\infty f(x), \qquad\qquad z = z_0 x^{-\beta},
\end{equation}
where $z_0=(R_4^+)^{1/2}$ is a universal constant, see Sec.~\ref{unratio}.
Because of Griffiths' analyticity, ${\cal F}(M)$ has also a regular
expansion in powers of $t$ for $|M|$ fixed. Therefore,
$F(z)$ has the large-$z$ expansion
\begin{equation}
F(z) = z^\delta \sum_{k=0} F^{\infty}_k z^{-k/\beta}.
\label{asyFz}
\end{equation}

\section{Parametric representations of the equation of state}
\label{preq}

\subsection{Polynomial approximation schemes}
\label{poly}

In order to obtain a representation of the equation of state that is
valid in the whole critical region, we need to extend analytically the
expansion (\ref{AZ}) to the low-temperature region $t<0$. For this
purpose, we use parametric representations that implement the expected
scaling and analytic properties.  We set \cite{parrep}
\begin{eqnarray}
M &=& m_0 R^\beta m(\theta) ,\nonumber \\
t &=& R(1-\theta^2), \nonumber \\
H &=& h_0 R^{\beta\delta}h(\theta), \label{parrep}
\end{eqnarray}
where $h_0$ and $m_0$ are normalization constants.  The variable $R$
is nonnegative and measures the distance from the critical point in
the $(t,H)$ plane, while the variable $\theta$ parametrizes the
displacement along the lines of constant $R$. The functions
$m(\theta)$ and $h(\theta)$ are odd and regular at $\theta=0$ and at
$\theta=1$.  The constants $m_0$ and $h_0$ can be chosen so that
$m(\theta)=\theta+O(\theta^3)$ and $h(\theta)=\theta+O(\theta^3)$.
The smallest positive zero $\theta_0$  of $h(\theta)$, which should satisfy
$\theta_0>1$, corresponds to the coexistence curve, i.e., to $T<T_c$
and $H\to 0$. 
The parametric representation satisfies the requirements of regularity
of the equation of state. Singularities can only appear at the
coexistence curve (due, for example, to the logarithms discussed in
Ref.\ \cite{PV-99}), i.e., for $\theta=\theta_0$.  
The mapping (\ref{parrep}) is not invertible when its Jacobian vanishes,
which occurs when
\begin{equation}
Y(\theta) \equiv (1-\theta^2)m'(\theta) + 2\beta\theta m(\theta)=0.
\label{Yfunc}
\end{equation}
Thus, parametric representations based on the mapping (\ref{parrep})
are acceptable only if $\theta_0<\theta_l$, where $\theta_l$ is the
smallest positive zero of the function $Y(\theta)$.  

\FIGURE[ht]{
\epsfig{file=fx.eps, width=12truecm} 
\caption{
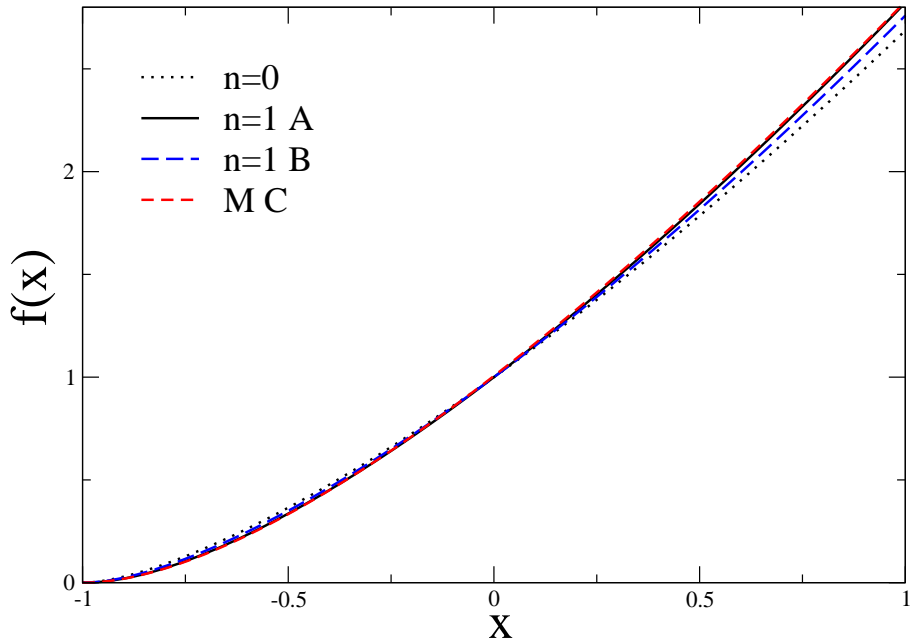
The scaling function $f(x)$ as obtained by the $n$=0, $n$=1 (A), and $n$=1 (B)
approximations. For comparison, we also show the Monte Carlo (MC) result 
of Ref.~\protect\cite{EM-00}, although it can be hardly distinguished
from the $n$=1 (A) curve.
}
\label{figfx}
}

\FIGURE[ht]{
\epsfig{file=Fz.eps, width=12truecm} 
\caption{
The scaling function $F(z)$.
}
\label{figFz}
}

The functions $m(\theta)$ and $h(\theta)$ are related to 
the scaling function $f(x)$ through
\begin{eqnarray}
&& x = {1 - \theta^2\over \theta_0^2 - 1} 
\left[ {m(\theta_0)\over m(\theta)}\right] ^{1/\beta} ,\label{fxmt}\\
&& f(x) = \left[ {m(\theta)\over m(1)}\right] ^{-\delta} {h(\theta)\over h(1)}.\nonumber
\end{eqnarray}
The asymptotic behavior (\ref{fxcc}) is reproduced by requiring that
\begin{equation}
h(\theta)\sim \left( \theta_0 - \theta\right)^2 
        \qquad\qquad{\rm for}\qquad \theta \rightarrow \theta_0.
\label{hcoex}
\end{equation}
Following Ref.\ \cite{CPRV-00}, we construct approximate polynomial
parametric representations that have the expected singular behavior at
the coexistence curve (Goldstone singularity) and match the known 
terms of the small-$z$ expansion of $F(z)$, cf. Eq.~(\ref{fzexp}).
We consider two distinct approximation schemes.  In the first one,
which we denote by (A), $h(\theta)$ is a fifth-order polynomial
with a double zero at $\theta_0$ and $m(\theta)$ is a polynomial of
order $(1+2n)$:
\begin{eqnarray}
{\rm scheme}\quad({\rm A}):\qquad\qquad 
&&m(\theta) = \theta 
    \Bigl(1 + \sum_{i=1}^n c_i \theta^{2i}\Bigr), \nonumber \\
&&h(\theta) = \theta \left( 1 - \theta^2/\theta_0^2 \right)^2. 
\label{scheme1}
\end{eqnarray}
In the second scheme, denoted by (B), we set 
\begin{eqnarray}
{\rm scheme}\quad({\rm B}):\qquad\qquad 
&&m(\theta) = \theta, \nonumber \\
&&h(\theta) = \theta 
    \left(1 - \theta^2/\theta_0^2 \right)^2
    \Bigl( 1 + \sum_{i=1}^n c_i \theta^{2i}\Bigr).
\label{scheme2}
\end{eqnarray}
Here $h(\theta)$ is a polynomial of order $5+2n$ with a double zero at
$\theta_0$.  For $n$=0 approximations (A) and (B) coincide.
It is worth noting that the $n$=0 approximation becomes
exact in $O(N)$ models for $N\to\infty$ \cite{BW-73,CPRV-00}.
In both schemes the parameter $\theta_0$ and the 
$n$ coefficients $c_{i}$ are
determined by matching the small-$z$ expansion of $F(z)$. 
The scaling function $F(z)$ can be written in terms of the parametric
representation as
\begin{eqnarray}
&&z = \rho \,m(\theta) \left( 1 - \theta^2\right)^{-\beta},
\label{Fzrel} \\
&&F(z(\theta)) = \rho \left( 1 - \theta^2 \right)^{-\beta\delta} h(\theta),
\nonumber
\end{eqnarray}
where $\rho$ is a normalization \cite{GZ-97,PV-r},
which is fixed by matching the expansion (\ref{fzexp}) to order $z^3$. 
Explicitly, we have 
\begin{equation}
\rho^2 = 6 \beta(\delta-1) - {12\over \theta_0^2} - 6 b c_1 ,
\end{equation}
where $b=1,-1$ respectively  for scheme (A) and (B). 
In both schemes, in order to fix $\theta_0$ and the $n$ coefficients $c_i$  
we need to know the values of $n+1$ coefficients $r_{2j}$, i.e., 
$r_6,...r_{6+2n}$.  This method has been already applied 
to the determine the critical equation of state of the
$XY$ \cite{CPRV-00,CHPRV-01} and Heisenberg \cite{CHPRV-02}
universality classes.

\subsection{Results}
\label{reseq}

\FIGURE[ht]{
\epsfig{file=Ey.eps, width=12truecm} 
\caption{
The scaling function $E(y)$.
}
\label{figEy}
}

In order to implement the above-presented approximation schemes,
we use the Monte Carlo estimates \cite{Hasenbusch-00}
$\nu=0.749(2)$ and $\eta=0.0365(10)$
for the critical exponents, and the field-theoretical 
estimates $r_6=1.79(2)$ and $r_8=0.2(4)$ which take into account
both fixed-dimension and $\epsilon$-expansion results, see
Sec. \ref{CES2}.
This will provide three different approximations:
$n$=0, $n$=1 (A), and $n$=1 (B). 
We find
\begin{equation}
\theta_0^2=2.795(40) \qquad n=0,
\end{equation}
and 
\begin{eqnarray}
&& \theta_0^2=2.949(150), \qquad c_1=-0.0225(200), \qquad n=1 \;({\rm A}),
\label{n1a}\\
&& \theta_0^2=2.373(200), \qquad c_1=\phantom{-}0.0650(300), \qquad n=1 \;({\rm B}),
\end{eqnarray}
where the number between parentheses indicates how much the coefficients vary
when the input parameters change by one error bar.
The relatively small value of $c_1$ 
in the $n$=1 approximations supports the effectiveness
of the approximation schemes.
In  Figs.~\ref{figfx}, \ref{figFz}, \ref{figEy}, and \ref{figDy} 
we show respectively 
the scaling functions $f(x)$, $F(z)$, $E(y)$, $D(y)$,
as obtained from the $n$=0, $n$=1 (A), and $n$=1 (B) approximations
for the central values of the input parameters.
The differences among the three approximations give
an indication of the size of the systematic error
of the approximation schemes.
The figures show that it is rather small,
suggesting that the $n=0,1$ representations
already provide good approximations of the critical equation of state.

\FIGURE[ht]{
\epsfig{file=Dy.eps, width=12truecm} 
\caption{
The scaling function $D(y)$.
}
\label{figDy}
}

In Fig.~\ref{figfx} we compare our results with the
scaling function $f(x)$
obtained in Ref.~\cite{EM-00} by interpolating 
Monte Carlo data for the O(4) spin model (\ref{spinmodel}).
The agreement is overall satisfactory.
The Monte Carlo determination of $f(x)$ turns out to be
hardly distinguishable from the $n$=1 (A) curve.
A more direct comparison with the Monte Carlo data of Ref.~\cite{EM-00}
is presented in Fig.~\ref{figeymc}, where we plot
$M/H^{1/\delta}$ versus $(J-J_c)/H^{1/\beta\delta}$
using $\delta=4.789(6)$, $\beta=0.388(1)$ and the raw data reported there.
To avoid finite-size and scaling corrections
we only consider the results for the largest lattices that are 
closest to the critical point, in
particular those satisfying  $H\le 0.01$ and $|J-J_c| < 0.05$.
The full line shown in Fig.~\ref{figeymc} is the 
$n$=1 (A) approximation, cf. Eqs.~(\ref{scheme1}) and
(\ref{n1a}), while the dotted lines show the uncertainty
due to the error on the corresponding input parameters.
The necessary normalization conditions
related to the amplitudes $B$ and $B_c$, cf. Eq.~(\ref{eyf}),
have been determined by fitting the data.
The agreement is good: most Monte Carlo data are within
the uncertainty of the $n$=1 (A) curve.

\TABLE[ht]{
\caption{
Results concerning the scaling functions obtained 
by using the $n$=0, $n$=1 (A), and $n$=1 (B) approximations.
Numbers marked with an asterisk are inputs, not predictions.  
}
\label{eqstdet0}
\begin{tabular}{ccccc}
\hline\hline
\multicolumn{1}{c}{}&
\multicolumn{1}{c}{$n=0$}&
\multicolumn{1}{c}{$n=1$ (A)}&
\multicolumn{1}{c}{$n=1$ (B)}&
\multicolumn{1}{c}{final estimates}\\
\hline \hline
$f^1_0$         & 1.44(1)    & 1.53(9)  & 1.48(3)    & 1.50(8) \\
$f^2_0$         & 0.29(1)    & 0.35(6)  & 0.32(2)    & 0.33(5) \\
$f^3_0$         & $-$0.060(4)&$-$0.08(1)& $-$0.07(1) & $-$0.08(2) \\
$f_0^\infty$    & 0.82(1)    & 0.92(10) & 0.87(3)    & 0.89(9) \\
$f_2^{\rm coex}$& 4.8(3)     & 2.2(1.3) & 3.3(6)     & 2.8(1.4) \\
$r_8$           & $-$0.3(1)  &$^*$0.2(4)& $^*$0.2(4) & \\
$r_{10}$        & 2.7(5)     & $-$4(4)  & $-$6(7)    & $-$5(6) \\
$F_0^\infty$    & 0.0236(3)  & 0.0241(5)& 0.0240(4)  & 0.0240(5) \\
$y_{\rm max}$   & 1.46(1)  & 1.35(10)  & 1.40(4)  & 1.38(9)  \\ 
$D(y_{\rm max})$& 0.3417(4)& 0.3429(11)& 0.3426(7)& 0.3427(10)\\ 
\hline\hline
\end{tabular}
}

In Table~\ref{eqstdet0} we report results concerning the behavior 
of the scaling functions $f(x)$ and $F(z)$ for $H=0$ and on the 
critical isotherm,  cf. Eqs. (\ref{expansionfx-xeq0}), 
(\ref{largexfx}), (\ref{fxcc}), (\protect\ref{AZ}), and (\ref{asyFz}).
We also report estimates of  $y_{\rm max}$ and $D(y_{\rm max})$,
where $y_{\rm max}$ is the value of $y$ corresponding to 
the maximum of the scaling function $D(y)$
defined in Eq.~(\ref{defDy-1}).
The displayed errors refer only to the uncertainty of the
input parameters and do not include the systematic error of the
procedure, which may be determined by comparing the results of the
various approximations.  
Our final estimates are reported in the Table~\ref{eqstdet0}.
They are obtained by taking the average of the results $n$=1 (A) and 
$n$=1 (B). The error is just indicative: it is
the sum of the uncertainty induced by the input parameters
(we take the average of the two errors)
and of half of the difference between the two approximations.
In most cases this procedure leads to an estimate that
includes the $n$=0 result within its error.
For comparison, we also report the estimates
$f_2^{\rm coex}=2.20(5)$,
$f_0^\infty=0.888(7)$, $F_0^\infty=0.018(4)$,
$y_{\rm max} = 1.33(5)$, and $D(y_{\rm max})=0.341(1)$, 
which can be obtained from the results of the fits of
the Monte Carlo data reported in Refs.~\cite{EM-00,EHMS-01}.\footnote{
The comparison of our results with the Monte Carlo 
estimates of Refs.~\cite{EM-00,EHMS-01} should be done
with caution. Monte Carlo results are subject to 
scaling corrections and finite-size effects. 
Moreover, Ref.~\cite{EM-00} used $\delta=4.86$ and $\beta=0.38$, while
we use the more precise estimates \cite{Hasenbusch-00} 
$\delta=4.789(6)$ and $\beta=0.388(1)$.
Our errors take into account the uncertainty on the critical exponents.
\label{footnt}}
Finally, we mention that the result for $r_{10}$ is
consistent with the estimate $r_{10}=9(17)$
obtained from the analysis of its $O(\epsilon^3)$ series
\cite{PV-effpot}.

\FIGURE[ht]{
\epsfig{file=eyca2.eps, width=12truecm} 
\caption{
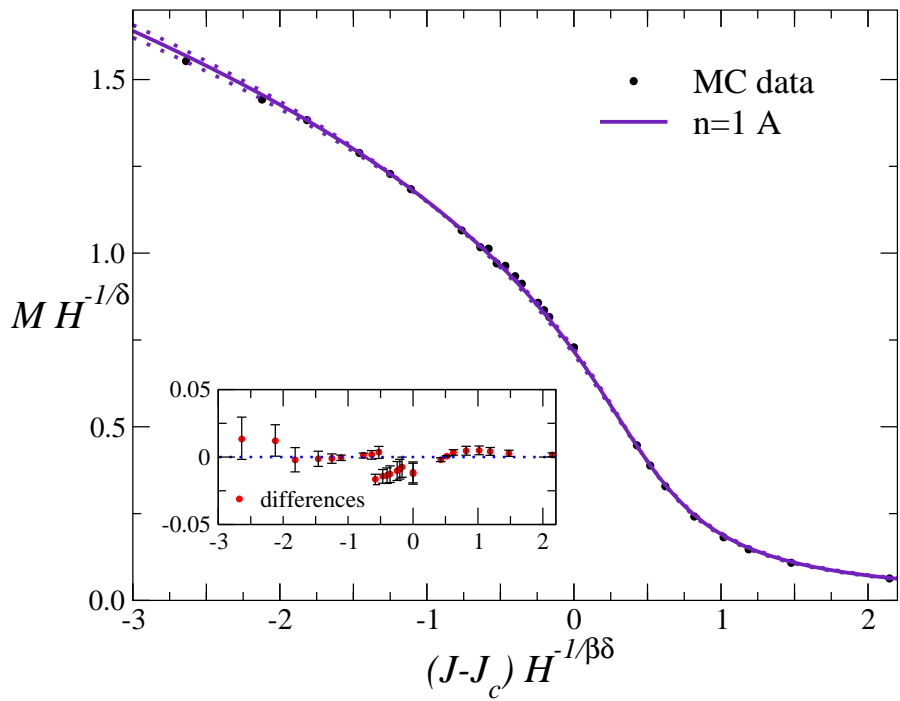
Monte Carlo data for $M/H^{1/\delta}$  versus $(J-J_c)/ H^{1/\beta\delta}$
from Ref.~\protect\cite{EM-00} compared with
the $n$=1 (A) approximation (full line).
The dotted lines show the uncertainty on the curve due to the errors on the 
input parameters.
The inset shows the differences between the Monte Carlo data and the $n$=1 (A) 
curve:
the errors are related to the uncertainty  on the theoretical curve; the
statistical errors of the Monte Carlo data are much smaller. 
}
\label{figeymc}
}

\section{Universal amplitude ratios}
\label{unratio}

\TABLE[ht]{
\caption{
Definitions of several universal amplitude ratios. 
}
\label{notationsur}
\begin{tabular}{ll}
\hline \hline
\multicolumn{2}{c}{\large{Universal Amplitude Ratios}}\\ 
\hline\hline
$U_0\equiv A^+/A^-$ 
& 
$R_c^+\equiv \alpha A^+C^+/B^2$ 
\\ [1.5mm]
$R_4^+\equiv - C_4^+B^2/(C^+)^3$ 
& 
$R_\chi\equiv C^+ B^{\delta-1}/B_c^\delta$ 
\\ [1.5mm]
$g_4\equiv -C_4^+/[ (C^+)^2 (f^+)^3]$ $\qquad\qquad$ 
&  
$Q^+ \equiv \alpha A^+ (f^+)^3$
\\ [1.5mm]
$R^+_\xi\equiv (Q^+)^{1/3}$
&
$Q_c \equiv B^2(f^+)^3/C^+$
\\ [1.5mm]
$ P_m \equiv { T_p^\beta B/B^c} $ & 
$ R_p \equiv { C^+/C_p} $ \\ [1.5mm]
\hline\hline
\end{tabular}
}

Universal amplitude ratios characterize the critical behavior
of thermodynamic quantities that do not depend on
the normalizations of the external (magnetic) field, of the order
parameter (magnetization), and of the temperature.  
From the scaling function $f(x)$ one may derive 
many universal amplitude ratios involving zero-momentum quantities, such as
the specific heat, the magnetic susceptibility, etc....
For example,
the universal ratio of the specific-heat amplitudes in the two phases
can be written as 
\begin{equation}
{A^+\over A^-} = {\varphi(\infty)\over \varphi(-1)}
\label{aratio}
\end{equation}
where, in the case of the O(4) universality class for which $-1<\alpha<0$,
\begin{equation}
\varphi(x) = 
{x |x|^{\alpha-2} f'(0) \over \alpha-1}  +
{|x|^{\alpha} f''(0)\over \alpha} - |x|^{\alpha-2} f(x) + 
\int_0^x d y\, |y|^{\alpha-2}
\left[ f'(y) - f'(0) - y f''(0) \right] .
\label{U0fx}
\end{equation} 
Several universal amplitude ratios 
can be expressed in terms of the amplitudes derived from
the singular behavior of the specific heat
\begin{equation}
C_H = A^{\pm} |t|^{-\alpha}+ b,
\end{equation}
where $b$ is a nonuniversal constant,
the magnetic susceptibility in the high-temperature phase
\begin{equation}
\chi = N C^{+} t^{-\gamma}
\end{equation}
with $N=4$,
the zero-momentum four-point connected correlation function in the
high-temperature phase
\begin{equation}
\chi_4 = {N(N+2)\over 3} C_4^+ t^{-\gamma-2\beta\delta}
\end{equation}
(again with $N=4$),
the second-moment correlation length in the high-temperature phase,
corresponding to the inverse mass in the zero-momentum renormalization
scheme, cf. Eqs.~(\ref{ren1}) and (\ref{ren2}),
\begin{equation}
\xi = f^{+} t^{-\nu},
\end{equation}
the spontaneous magnetization on the coexistence curve, cf. Eqs. (\ref{bc})
and (\ref{bt}).
We also consider amplitudes along the crossover line $t_{\rm max}(H)$, 
that is defined as the reduced temperature where
the longitudinal magnetic susceptibility $\chi_L(t,H)=\partial M / \partial H$ 
has a maximum at $H$ fixed, i.e.
\begin{eqnarray}
&&t_{\rm max}(H) = T_p H^{1/(\gamma+\beta)},\\
&&\chi_L(t_{\rm max},H)= C_p t_{\rm max}^{-\gamma}.
\end{eqnarray}
In Table \ref{notationsur} we report the definitions of several
universal amplitude ratios that have been considered in the literature.
Their expressions in terms of the 
functions $m(\theta)$ and $h(\theta)$ can be found in Ref.~\cite{CHPRV-02}.
Note the following relations 
\begin{eqnarray}
&&f_0^\infty = R_\chi^{-1}, \qquad f_1^\infty= {R_4^+\over 6 R_\chi},\\
&&y_{\rm max} = P_m^{1/\beta}, \qquad
D(y_{\rm max}) = R_p^{-1} P_m^{1-\delta} R_\chi.
\end{eqnarray}
where $f_0^\infty$ and $f_1^\infty$  are related to the 
large-$x$ behavior of $f(x)$, cf. Eq.~(\ref{largexfx}), 
and $y_{\rm max}$ is the value of $y$ where the scaling
function $D(y)$ takes its maximum.

In Table \ref{univratio} we report the estimates of 
several universal amplitude ratios,
as derived by using the approximations
$n$=0, $n$=1 (A), and $n$=1 (B).
Again, the displayed errors refer only to the uncertainty of the
input parameters. Our final estimates 
are determined as in the case of the quantities reported in 
Table~\ref{eqstdet0}.
They can be again compared with the Monte Carlo results reported in 
Refs.~\cite{EM-00,EHMS-01}:
$R_\chi=1.126(9)$, $R_4^+=8.6(9)$
and $P_m=1.11(2)$. There is good agreement, keeping again into 
account that Refs.~\cite{EM-00,EHMS-01} used different values of the 
critical exponents (see footnote \ref{footnt}).
Moreover, using  Eq.~(\ref{aratio}) and  the approximate interpolation formula 
of Refs.~\cite{EM-00,EHMS-01} we obtain
$U_0=2.0(2)$, where the error is obtained 
by varying the curve parameters within their error. 
We also mention the estimates \cite{KV-01}
$U_0\approx 2.044$ and
$R_c^+ \approx 0.263$, obtained by an analysis of three-loop
series in the minimal-subtraction scheme without
$\epsilon$ expansion.
They are in good agreement with those presented in Table \ref{univratio}.

Universal amplitude ratios 
involving the correlation-length amplitude $f^+$, such as 
$R_\xi^+$ and $Q_c$, can be obtained by using the
estimate of $g_4$. If we take \cite{GZ-98} $g_4=17.30(6)$,  we obtain
\begin{eqnarray}
&&R_\xi^+ = \left( {R_c^+ R_4^+ \over g_4} \right)^{1/3}= 0.490(4), \\
&&Q_c={R_4^+\over g_4}= 0.44(2). 
\end{eqnarray}

\TABLE[ht]{
\caption{
Estimates of several universal amplitude ratios.
}
\label{univratio}
\begin{tabular}{ccccc}
\hline\hline
\multicolumn{1}{c}{}&
\multicolumn{1}{c}{$n=0$}&
\multicolumn{1}{c}{$n=1$ (A)}&
\multicolumn{1}{c}{$n=1$ (B)}&
\multicolumn{1}{c}{final estimates}\\
\hline \hline
$U_0$           & 2.03(3)  & 1.89(11)  & 1.93(6)  & 1.91(10) \\
$R_c^+$         & 0.250(5) & 0.274(25) & 0.263(9) & 0.27(2) \\
$R_4^+$         & 8.04(7)  & 7.5(5)    & 7.7(2)   & 7.6(4)  \\
$R_\chi$        & 1.22(2)  & 1.09(12)  & 1.15(4)  & 1.12(11) \\
$P_m$           & 1.159(3) & 1.12(3)   & 1.140(11)& 1.13(2) \\
$R_p$           & 2.049(3) & 2.041(7)  & 2.042(6) & 2.042(7) \\
$R_c^+\,R_4^+$  & 2.01(4)  & 2.05(5)   & 2.03(3)  & 2.04(5) \\
\hline\hline
\end{tabular}
}

\acknowledgments{ We thank Luigi Del Debbio for useful
and interesting discussions.}

\end{document}